\documentclass[utf8]{frontiersSCNS} 

\usepackage{url,hyperref,lineno,microtype,subcaption}
\usepackage[onehalfspacing]{setspace}
\usepackage{natbib}

\newcommand{\hii}{\mbox{H\,\textsc{ii}}}
\newcommand{\hi}{\mbox{H\,\textsc{i}}}

\def\keyFont{\fontsize{8}{11}\helveticabold }
\def\firstAuthorLast{Hou L.G.} 
\def\Authors{L.G. Hou\,$^{1,2}$}
\def\Address{$^{1}$National Astronomical Observatories, CAS, Jia-20 Datun Road, Chaoyang District, Beijing 100101, PR China\\
  $^{2}$CAS Key Laboratory of FAST, National Astronomical Observatories, Chinese Academy of Sciences}
\def\corrAuthor{L.G. Hou}

\def\corrEmail{lghou@nao.cas.cn}

\begin{document}
\onecolumn
\firstpage{1}

\title[Nearby spiral structure]{The spiral structure in the Solar
  neighborhood}

\author[\firstAuthorLast ]{\Authors} 
\address{} 
\correspondance{} 

\extraAuth{}%

\maketitle

\begin{abstract}

The spiral structure in the Solar neighborhood is an important issue
in astronomy. In the past few years, there is significant progress in
observation. The distances for a large number of good spiral tracers,
i.e. giant molecular clouds, high-mass star-formation region masers,
$\hii$ regions, O-type stars and young open clusters, have been
accurately estimated, making it possible to depict the detailed
properties of nearby spiral arms. In this work, we first give an
overview about the research status for the Galaxy's spiral structure
based on different types of tracers. Then the objects with distance
uncertainties better than 15\% and $<$0.5~kpc are collected and
combined together to depict the spiral structure in the Solar
neighborhood. Five segments related with the Perseus, Local,
Sagittarius-Carina, Scutum-Centaurus and Norma Arms are traced. With
the large dataset, the parameters of the nearby arm segments are
fitted and updated.  Besides the dominant spiral arms, some
substructures probably related to arm spurs or feathers are also
noticed and discussed.

\tiny \keyFont{ \section{Keywords:} $\hii$ regions -- Galaxy:
  structure -- stars: early-type -- stars: massive -- masers}
\end{abstract}

\section{Introduction}
\label{sec:intro}

As observers deeply embedded in the Galactic disk, mapping the spiral
structure of the Milky Way and understanding its formation and
evolution have long been difficult issues in astronomy.
Superpositions of multiple structures along the same observed line of
sight have to be solved to trace the distribution of matters in our
Galaxy.
Additionally, the widespread dust in the interstellar medium
causes extinction, making the situation more complex.
However, because we live in the Milky Way, the positions and
kinematics for a large number of objects could be measured with high
accuracy, making the Milky Way as the only galaxy in the universe that
we can investigate in detail.

Spiral structure is one of the fundamental characteristics of the
Milky Way. It has considerable influences on some other research
fields, such as the kinematics of nearby stars
\citep[e.g.][]{wsb13,hbb+19,tfh+21}, the Galactic electron-density
distribution \citep[][Han et al. 2021]{tc93,ymw17}, the Galactic dust
distribution and extinction map \citep[][]{ds01,hba20}, and the
large-scale magnetic field of the Milky Way \citep[e.g.][]{han17}.
There have been quite a few reviews about the global properties of
Galaxy's spiral structure, \citet{fc10}, \citet{xhw18} and
\citet{sz20} reviewed previous efforts. Although, disagreements remain
in some details, a general consensus that a global spiral pattern
exists in the Galactic disk has been achieved.
In this work, we focus on the Solar neighborhood, where the spiral
structure can be better understood, because the distances of a large
number of nearby objects can be measured accurately. Significant
progress has been made in the past few years by taking advantage of
the astrometry measurements in radio to optical bands, which enables
us to reliably delineate the nearby spiral structure in unprecedented
detail.

\citet[][]{mor52,mor53} first delineated three arm segments in the
Solar neighborhood with a sample of aggregates of high-luminosity O-A
stars in the 1950s. The three segments are now known to be related to
the Sagittarius-Carina, Local and Perseus Arms. At that time, these
structures were also studied by using different methods
\citep[e.g.][]{th56,bok64,bok70,gg71}, such as, mapping the
distribution of $\hii$ regions \citep[][]{cgg70}, analyzing multiple
structures shown in the $\hi$ 21~cm line surveys \citep[][]{vmo54} or
implied in the observational data of other interstellar absorption
lines toward background stars \citep[][]{munch53}.
In the 1970s, \citet{gg76} proposed the famous model of the Galaxy's
spiral structure consisting of four major spiral arm segments. The Sun
was placed in the inter-arm region between the Perseus Arm and the
Sagittarius Arm.
Then, the picture of Galaxy's spiral structure was extended by taking
advantage of more observational data of different types of spiral
tracers e.g., $\hii$ regions \citep[e.g.][]{dwbw80,ch87,rus03,pal04},
molecular clouds \citep[e.g.][]{cohen80,cohen85,hhs09,lra11}, neutral
atomic gas \citep[][]{sim70,bur73,lev06,koo17}, high-mass
star-formation region (HMSFR) masers~\citep[e.g.][]{xrzm06,rmb+09}, OB
stars \citep[e.g.][]{mil72,sf74,dhb+99,wri20}, open clusters
\citep[e.g.][]{beck64,beck70,jan88,dl05} and cepheids
\citep[e.g.][]{fer68,mtl09}.
These great efforts enhanced our understanding of the global
properties of Galaxy's spiral structure.
For the spiral structure in the Solar neighborhood, \citet[][]{xlr+13}
first found that many HMSFR masers with accurate VLBI parallax
measurements \citep[][]{xrzm06} are situated in the Local Arm,
indicating that the Local Arm is probably a major arm segment, rather
than an inter-arm spur or a branch as has been suggested for a long
time.
The existence of the Local Arm also challenges the formation mechanism
of Galaxy's spiral structure \citep[][]{xrd+16}, since it would be
difficult to explain its existence according to the standard
density-wave theory \citep[][]{yuan69,shu16}, owing to the narrow
space between the Sagittarius Arm and Perseus Arm.

In the past few years, there have been significant progress in
observations by taking advantage of the VLBI observations in radio
band \citep[][]{rmb+19,VERA} and {\it Gaia} astrometry measurements in
optical band \citep[][]{gaia}.
Accurate parallaxes and proper motions have been obtained for a large
number of HMSFR masers \citep[e.g.][]{rmb+19}, $\hii$ regions
(e.g. Hou et al. 2021), O-type stars \citep[e.g.][]{xhb+21}, young
open clusters \citep[OCs, e.g.][]{dias21} and evolved stars
\citep[e.g.][]{kgd+20}. Additionally, many giant molecular clouds
(GMCs) have had accurately determined distances based on the
multi-wavelength survey data from optical to infrared bands or the
astrometric data of foreground/background stars
\citep[][]{yys+19,cly+20}.
By combining the available data of different types of tracers
together, it is now possible to reliably map the detailed spiral
structure within about 5~kpc of the Sun.

In this work, we first give an overview about the observational status
for each type of spiral tracer, the available dataset of these objects
with accurate distances are collected. Then, we combine them together
to give a detailed description of the spiral structure in the Solar
neighborhood. Conclusions and discussions follow in the last section.

\section{An overview of spiral tracers}

Young objects (HMSFR masers, $\hii$ regions, massive OB stars and
young open clusters etc.) and GMCs are known as good tracers for the
Galaxy's spiral structure (hereafter gas arms).
In addition, the spiral structure is imprinted by the distribution of
old and evolved stars (hereafter stellar arms).
The gas arms and stellar arms in a galaxy are not necessarily
coincident with each other. Discovering a large number of spiral
tracers widely spread through the Galactic disk, and measuring their
distances as accurately as possible are the key to settle the disputes
on the spiral structure of our Milky Way Galaxy.
In the following, the observational status about GMCs, HMSFR masers,
$\hii$ regions, OB stars, young OCs and evolved stars are discussed,
respectively.

\subsection{Giant molecular clouds}

Giant molecular clouds are the vast assemblies of molecular gas with
masses from $\sim$10$^3M_\odot$ to $\sim$10$^7M_\odot$
\citep[e.g.][]{murr11}. They are believed to primarily form in spiral
arms \citep[][]{db14} and are the nurseries of most young stars in a
galaxy.
In the Milky Way, GMCs have long been known as good tracers of spiral
arms \citep[e.g.,][]{mdt+86,gcbt88,hh14}.
From the wealthy dataset of Galactic CO surveys \citep[see][ for a
  review]{hd15}, a large number of isolated molecular clouds have been
identified by different methods
\citep[e.g.][]{gbn+14,rgb+16,mme17,yys+20,dua21}. For the majority of
them, only kinematic distances are known, which depend on the adopted
Galaxy rotation curve, the solution of the kinematic distance
ambiguity, and deviation from the hypothetic circular rotation. For
instance, \citet{dua21} compiled a large catalogue of 10,663 molecular
clouds in the inner Galaxy. They estimated the kinematic distances for
10,300 clouds after solving the distance ambiguities through different
methods.
In addition to determining the distances of molecular clouds and then
mapping their distribution in the Galactic disk, the other two methods
have been used to reveal the Galaxy spiral structure with CO sruveys:
(1) Deconvolution of the survey data cube \citep[e.g.][]{poh08,ns16};
(2) Modelling the observed longitude-velocity ($l-v$) maps of CO
\citep[e.g.][]{bis03,rc08,bsw10,pet14,pet15,li21}. A detailed review
about these two methods can be found in \citet{xhw18}. As discussed in
\citet{xhw18}, although there have been many efforts, the spiral
structure traced by molecular gas is still unclear, primarily due to
the large uncertainties of distances.
There have been noticeable progress in the past few years, as accurate
distances were estimated for a large number of molecular clouds.

With the CO data of the Milky Way Imaging Scroll Painting
survey\footnote{http://www.radioast.nsdc.cn/mwisp.php}, the {\it Gaia}
DR2 parallax and {\it G}-band extinction \citep[][]{dr2},
\citet{yys+19} proposed a background-eliminated extinction-parallax
method to estimate the distances of molecular clouds. The distance
uncertainties of 11 clouds are $\lesssim10\%$. With the same method,
\citet{yys+20} determined the distances for 28 local molecular clouds
($d<1.5$~kpc, here $d$ is the distance to the Sun) in the first
Galactic quadrant. The distances for 76 molecular clouds were measured
in the second Galactic quadrant by \citet{yys+21}.
Based on a sample of over 32~million stars with colour excesses and
{\it Gaia} distances, \citet{chy+19} constructed new three-dimensional
dust reddening maps of the Milky Way. With the maps and the sample of
stars, \citet[][]{cly+20} identified 567 dust/molecular clouds, and
estimated their distances by using a dust model fitting algorithm. The
typical distance uncertainty is less than 5\%. These clouds are within
$\sim3$~kpc of the Sun.
Based on the near-infrared photometry data from the Two Micron All Sky
Survey and the Vista Variables in the Via Lactea Survey,
\citet[][]{chh19} tracked the extinction of red clump stars versus
distance profiles of the sightlines towards a sample of molecular
clouds from \citet[][]{rgb+16}. Distances of 169 GMCs in the fourth
Galactic quadrant were obtained.

We collected the data of GMCs from above references. 475 GMCs with
masses $>10^4M_{\odot}$ were obtained. To reliably depict the local
spiral structure, only the GMCs with distance uncertainties better
than 15\% were adopted. For the distant clouds, we also required that
their distance uncertainties are $<$0.5~kpc. In total, 427 GMCs with
masses from $10^4 M_{\odot}$ to 2.45$\times10^6 M_{\odot}$ were left.
Their projected distributions on the Galactic disk are shown in
Fig.~\ref{dis}a.
According to the trigonometric parallax data of HMSFR masers (see
Sect.~\ref{sec:maser}), \citet{rmb+19} obtained an updated model of
Galaxy's spiral arms, which is plotted in Fig.~\ref{dis} to make a
comparison with the GMC distribution.
It is shown that most of these GMCs are distributed within about 3~kpc
of the Sun, in the Perseus, Local and Sagittarius-Carina Arms. Some
distant GMCs in the fourth Galactic quadrant are probably associated
with the Centaurus Arm and the Norma Arm.
Along spiral arms, the distribution of GMCs presents some
substructures, especially in the regions within about 2~kpc of the
Sun.
The accuracies of distances ensure that they are probably true
features, but their nature (arm spurs or feathers) and properties have
not been well studied.

\subsection{High-mass star-forming region masers}
\label{sec:maser}

\begin{figure*}
  \includegraphics[width=0.41\textwidth]{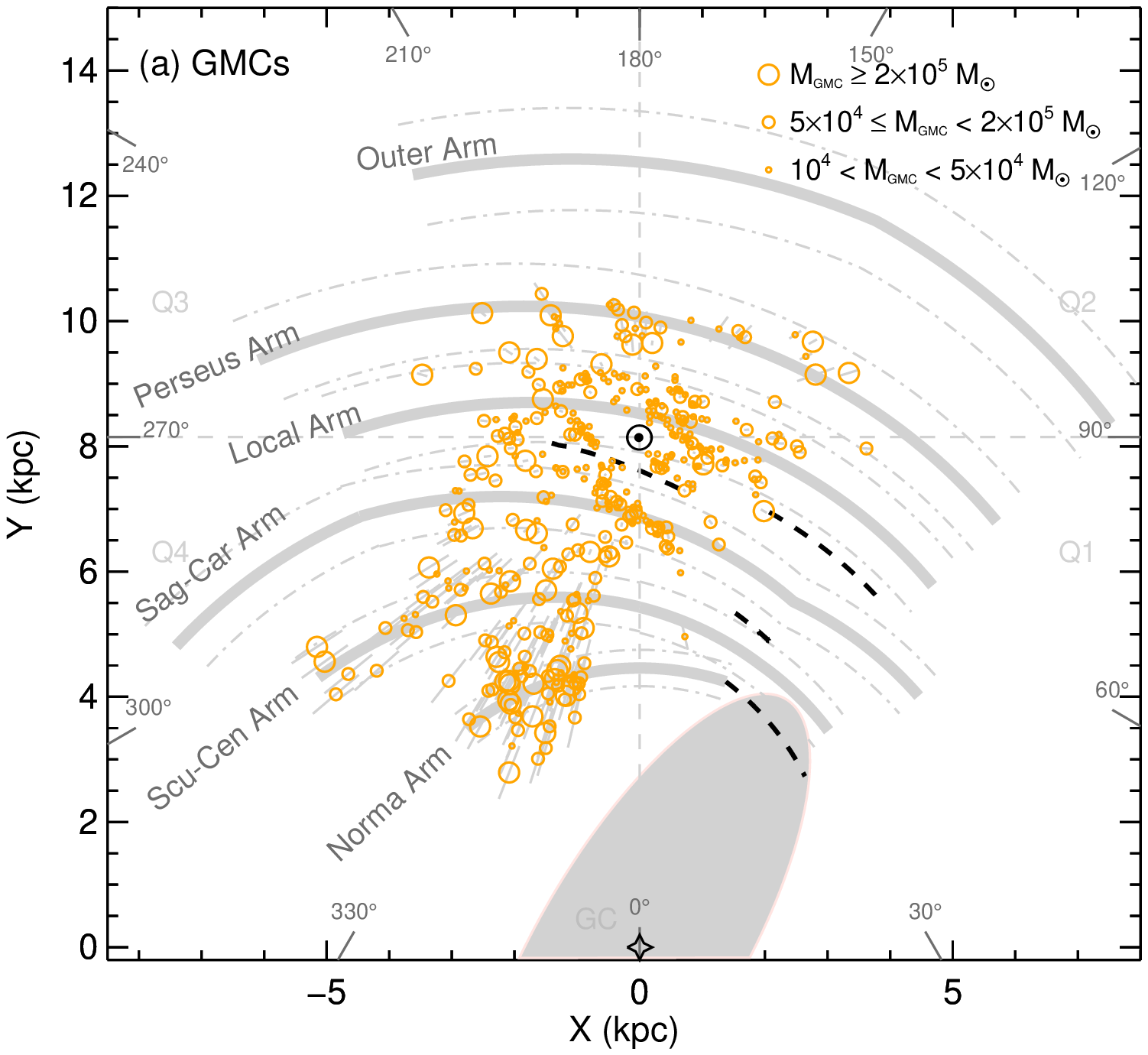}  
  \centering \includegraphics[width=0.41\textwidth]{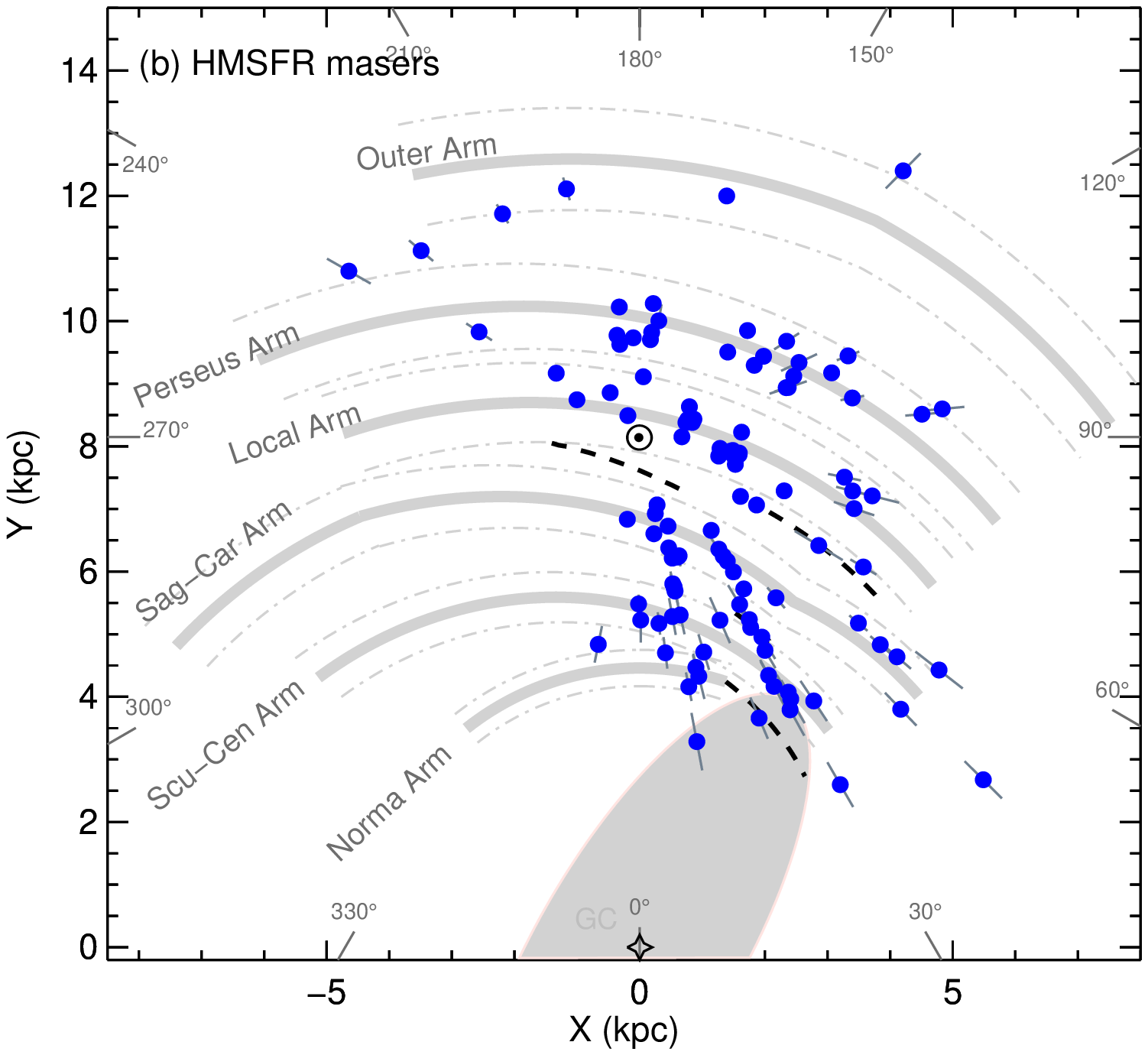}\\
  \includegraphics[width=0.41\textwidth]{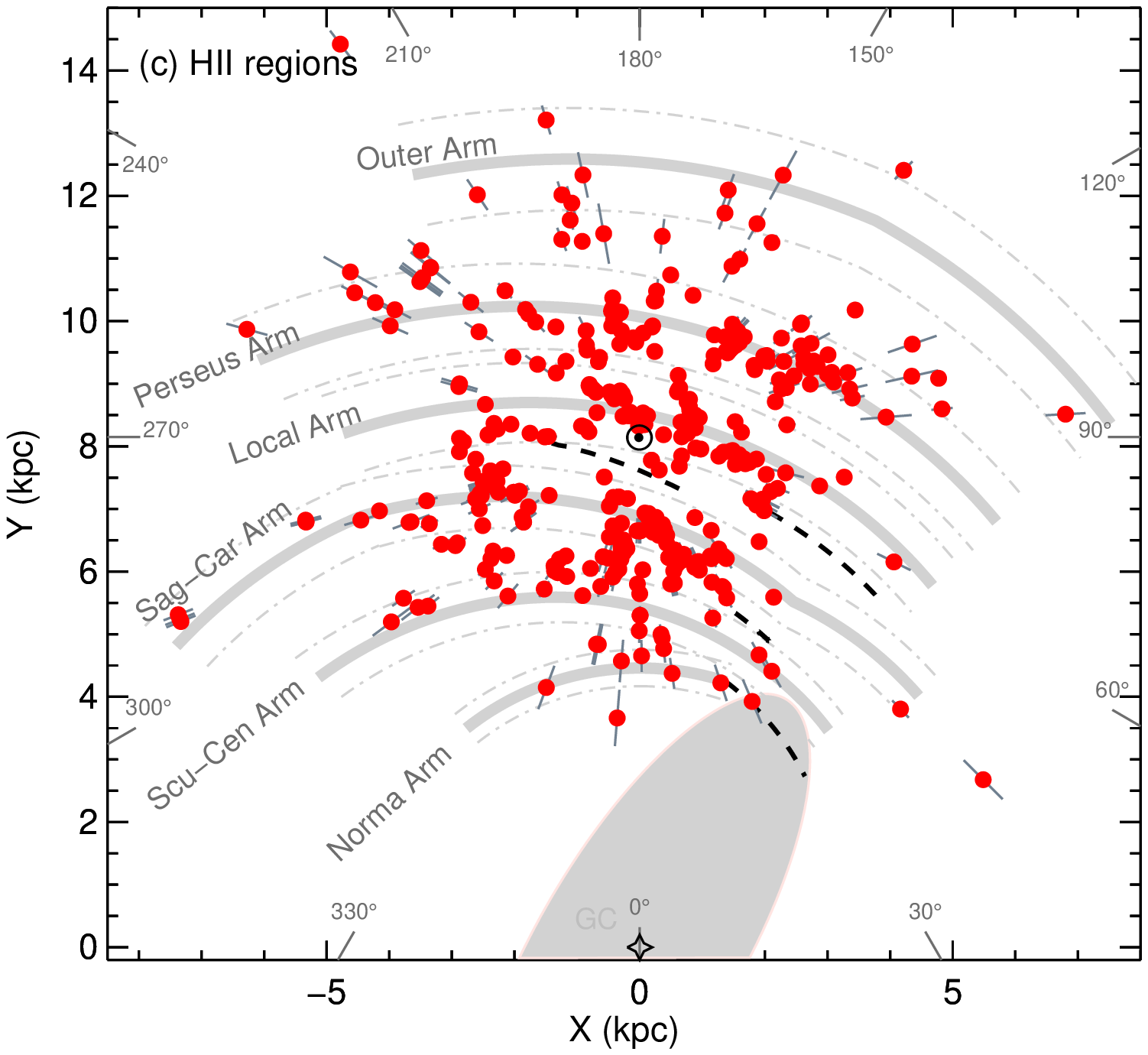}
  \includegraphics[width=0.41\textwidth]{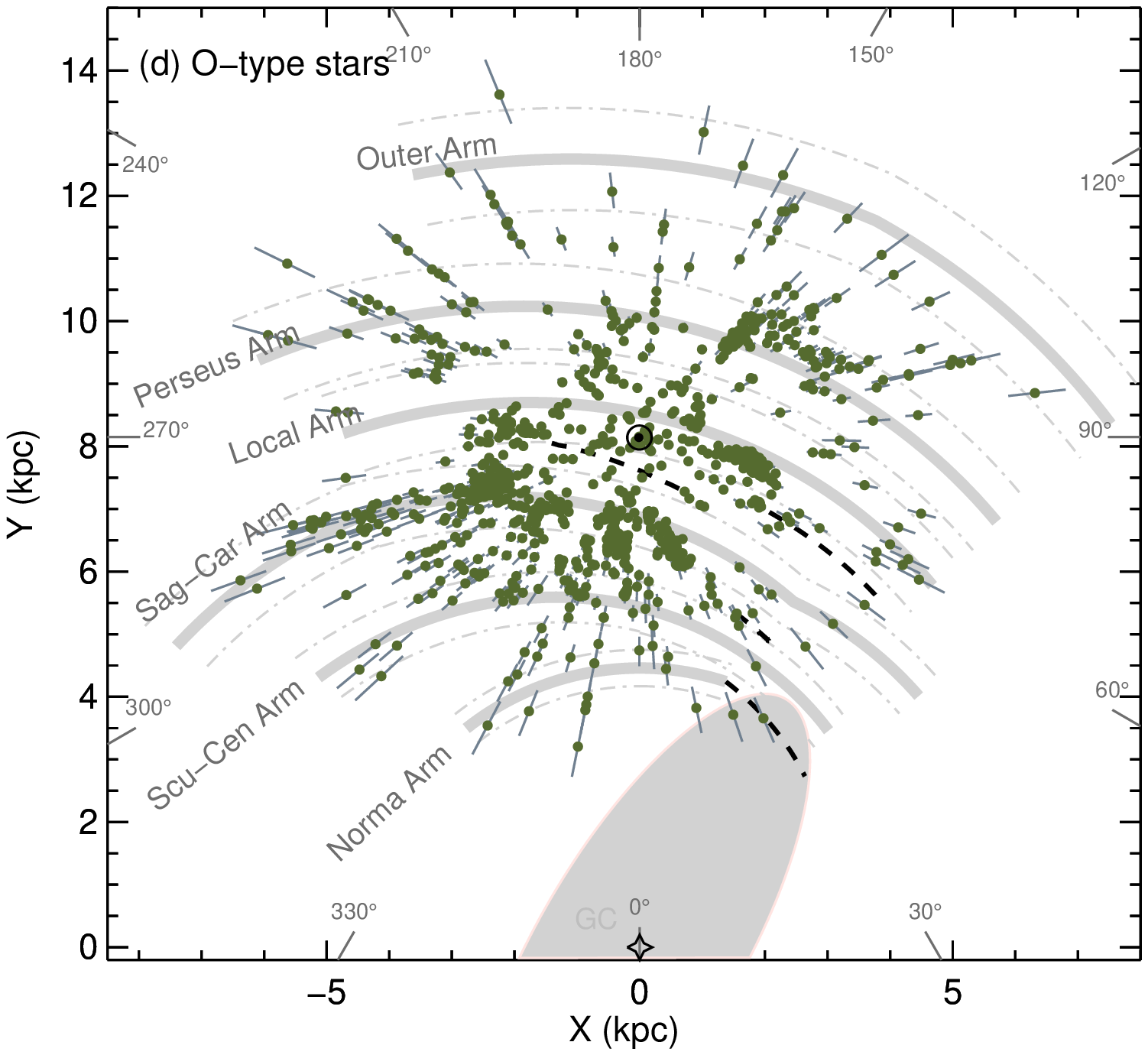} \\
      \includegraphics[width=0.41\textwidth]{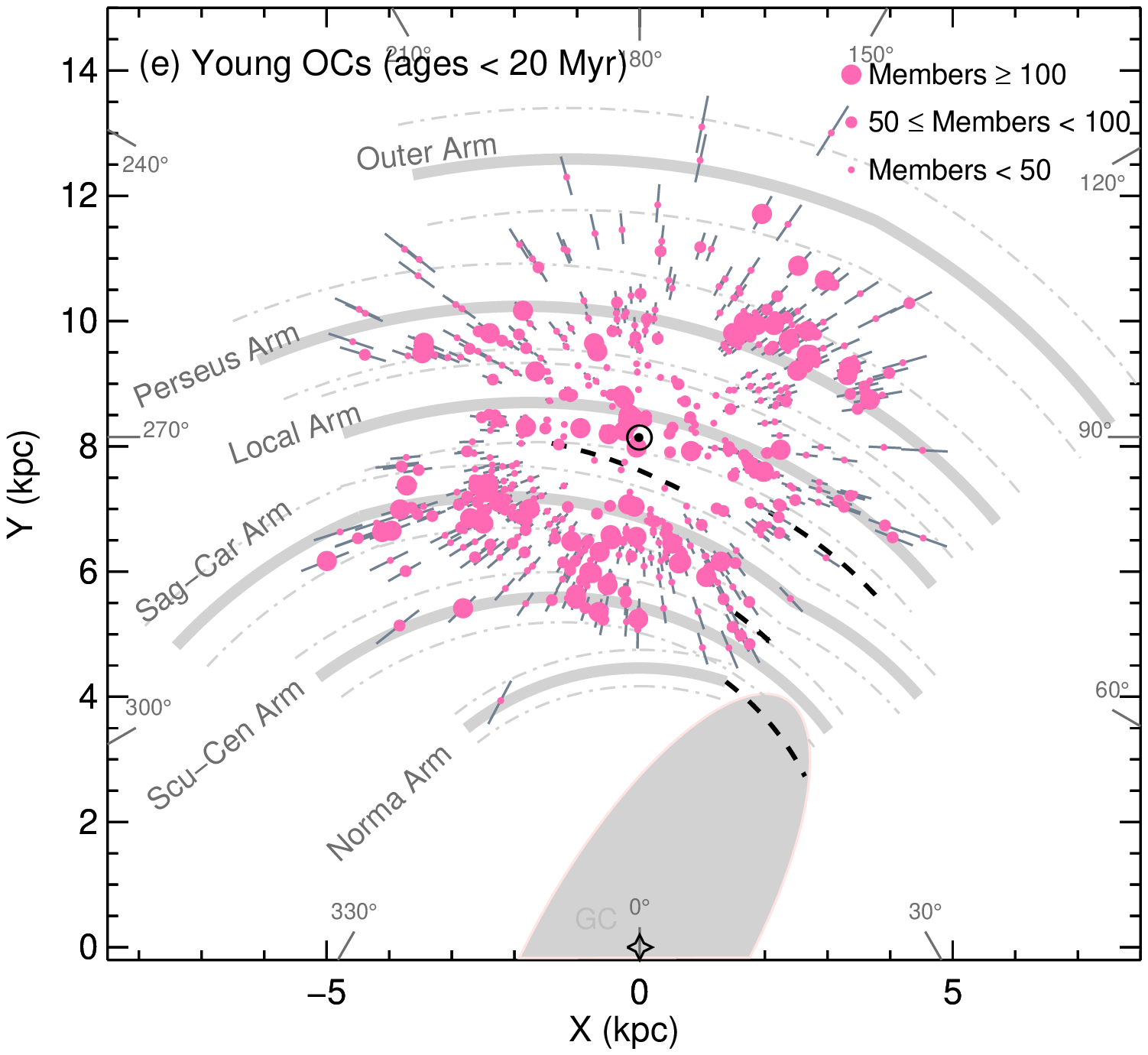} 
  \caption{Distributions of (a) GMCs, (b) HMSFR masers, (c) $\hii$
    regions, (d) O-type stars and (e) young OCs in the Solar
    neighborhood. All of the plotted objects have distance
    uncertainties better than 15\% and $<$0.5~kpc as determined by
    trigonometric or photometric method. The symbol sizes of GMCs are
    proportional to their masses. For HMSFR masers, $\hii$ regions and
    O-type stars, an equal size of the dots is adopted. The symbol
    sizes of young OCs are proportional to the number of cluster
    member stars. The position uncertainty for each data point is
    shown by an underlying gray line segment. The thick gray curved
    lines indicate the best fitted spiral arm model given by
    \citet{rmb+19}, the dotted lines denote the arm widths enclosing
    90\% of the HMSFR masers. The black dashed lines indicate the four
    spurs or spur-like structures proposed in literature (see
    Sect.~\ref{sec:spur}). In each plot: the segments of spiral arms
    are labelled; the Sun is at (0.0, 8.15)~kpc \citep[][]{rmb+19},
    while the Galactic center is at (0.0, 0.0)~kpc; the Galactic long
    bar is indicated by a shaded ellipse \citep[][]{weg15}.}
\label{dis}
\end{figure*}

The early stage of massive star formation is accompanied by the maser
emission from molecular species such as OH, CH$_3$OH and H$_2$O
\citep[][]{fish07}. The maser spots are compact and bright, hence are
optimal targets for radio interferometric observations. The
trigonometric annual parallax measurement with Very Long Baseline
Interferometry (VLBI) is the most accurate method for deriving the
distances of astronomical objects.
In 2006, a pioneer research on measuring the trigonometric parallax of
molecular masers was made by \citet[][]{xrzm06}, who found a distance
of W3(OH) in the Perseus Arm to be 1.95$\pm$0.04~kpc. This work opened
a new era to accurately reveal the Galaxy's spiral structure through
VLBI measurements.
Since then, nearly 200 HMSFRs have had measured trigonometric
parallaxes (typical accuracy is about $\pm$ 0.02 mas) and proper
motions, primarily by the Bar And Spiral Structure Legacy (BeSSeL)
Survey using the VLBA~\citep[][]{rmb+19} and the Japanese VLBI
Exploration of Radio Astrometry (VERA) project \citep[][]{VERA}. Some
sources were observed by the European VLBI Network and the Australian
Long Baseline Array~\citep[e.g.][]{rbr+10,ker+17}.
Based on the data of 199 HMSFRs, parameters of spiral arms in about
one third of the Galactic disk were updated by \citet[][]{rmb+19},
which is adopted in Fig.~\ref{dis} to make a comparison with the data
distribution.

To depict the local spiral arms with high confidence, only the HMSFRs
\citep[][]{rmb+19,VERA} with uncertainties of trigonometric distances
better than 15\% were kept as for GMCs. For the distant sources, we
also required that their distance uncertainties are $<$0.5~kpc. Then,
111 HMSFRs remain. Their distribution is given in
Fig.~\ref{dis}b. These HMSFRs are located in six arm segments,
i.e. the Outer, Perseus, Local, Sagittarius, Scutum and Norma
Arms. Some of the HMSFRs are probably related with spur-like
structures in the inter-arm regions \citep[][]{rmb+19}. A prominent
one is the spur branching the Sagittarius Arm and the Local Arm near
$l\sim50^\circ$ (see Fig.~\ref{dis}), which is firstly identified by
\citet{xrd+16}.

As shown in Fig.~\ref{dis}b, the HMSFRs with parallax measurements are
distributed in the regions covering about one-third of the entire
Galactic disk. There is a lack of observational data for many Galaxy
areas, especially in the longitude range of
$\sim240^\circ-360^\circ$. Other spiral tracers (e.g. GMCs, $\hii$
regions, massive OB stars, and young OCs) could be good complementary
data, which can help us to better depict the properties of spiral arms
in the Solar neighborhood.

\subsection{$\hii$ regions}

$\hii$ regions are the regions of ionized gas surrounding recently
formed O- or early B-type stars or star clusters. They have been
widely detected in the Galactic disk, from the Galactic center (GC)
region \citep[][]{ssn+19} to the far outer Galaxy
\citep[Galactocentric distances $>$16~kpc,][]{aaj+15}. As indicators
of early evolutionary stage of massive stars, $\hii$ regions have long
been known as one of the primary tracers of spiral arms, and helped us
to construct the commonly used picture of Galaxy's spiral structure
\citep[e.g.,][]{gg76,ch87,rus03,hhs09,hh14}.
However, for the majority of known $\hii$ regions, only kinematic
distances are available, which sometimes have large uncertainties. It
is currently the major obstacle for using $\hii$ regions to accurately
delineate the spiral arms.
There are about 400 $\hii$ regions with measured spectra-photometric
\citep[][]{rus03,mdf+11,fb15} or trigonometric          distances
\citep[e.g.][]{xrzm06,honm12}. They are distributed within about 5~kpc
of the Sun \citep[][]{hh14}. The sample size and the accuracies of
distances have yet to be improved.

In a recent work (Hou et al. 2021), we carried out a cross-match
between the WISE $\hii$ regions \citep[][]{wise} and the known O- or
early B-type stars \citep[][]{chh19,xhb+21}. The ionizing stars of 315
$\hii$ regions were identified. The trigonometric parallaxes for these
OB stars from the {\it Gaia} Early Data Release 3 \citep[{\it Gaia}
  EDR3,][]{edr3} were used to estimate the distances of $\hii$
regions. In combination with the $\hii$ regions with known
spectra-photometric/trigonometric distances, we obtain a sample of 448
$\hii$ regions with accurately determined distances, i.e. distance
uncertainties are better than 15\% and less than 0.5~kpc. Their
distribution in the Galactic disk resembles GMCs and HMSFR masers as
presented in Fig.~\ref{dis}c.

These $\hii$ regions are scattered primarily in the Perseus, Local,
Sagittarius-Carina and Scutum-Centaurus Arms.
The data distribution in the Local Arm and Centaurus Arm deviate from
the modelled spiral arms given by \citet{rmb+19} in the 3rd and 4th
Galactic quadrants, where the HMSFRs with trigonometric measurements
are largely absent.
Hence, the extension of the Local Arm and the position of the
Centaurus Arm given by \citet{rmb+19} need to be updated.
In these arm segments, we also noticed that the $\hii$ regions are not
uniformly distributed, but present some substructures.
The Sagittarius-Carina Arm is not continuous in the direction of
$l\sim$315$^\circ-340^\circ$. Similar property is found for the
Perseus Arm in the longitude range of $\sim$150$^\circ-160^\circ$.
These properties are consistent with the features illustrated by GMCs
(Fig.~\ref{dis}a).
Although the number of $\hii$ regions with accurate distances have
been increased largely, there is still a lack of accurate distances
for many Galactic $\hii$ regions. From {\it Gaia} EDR3 or their future
data release, it is expected to identify more ionising stars and
determine the distances of $\hii$ regions as accurately as possible.

\subsection{OB stars}

The massive and bright O- and early B-type (OB) stars are born in
dense molecular clouds. Many of them are not randomly distributed, but
concentrated in loose groups, named as ``aggregates''
\citep[][]{mor53} or OB associations \citep[][]{dhb+99,wri20}. In the
early 1950s, substantial progress on tracing the arm segments in the
Solar neighborhood was first made by \citet{mor52,mor53}. Three
segments of spiral arms (i.e., the Sagittarius, Local and Perseus
Arms) appeared in the distribution of their twenty-seven aggregates of
O-A stars and additional eight distant stars.
After that, some follow-up studies identified more OB stars or OB
associations, and determined their spectra-photometric distances
\citep[e.g.,][]{wal71,mil72,sf74,rr00}. However, the picture of Morgan
did not expand significantly.
With the {\it Hipparcos} catalog, \citet[][]{dhb+99} estimated the
trigonometric distances for some OB associations within $\sim$1~kpc
from the Sun, primarily in the Local Arm. At present, the known OB
associations are limited within about $\sim$2~kpc from the Sun
\citep[][]{wri20}. To extend the nearby arm segments traced by OB
stars or OB associations, higher accuracies of astrometry measurements
than the {\it Hipparcos} \citep[typical parallax error
  $\sim$1~mas,][]{dhb+99} are needed.

The {\it Gaia} satellite \citep[][]{gaia}, launched in 2013, will
ultimately achieve parallax accuracy comparable to that of VLBI for
approximately $10^9$ stars.
Many OB stars with accurate distances can be derived from {\it Gaia}.
By taking advantage of the {\it Gaia} data release 2, \citet{xbr+18}
depicted the spiral structure within $\sim$3 kpc of the Sun. About
2,800 O-B2 stars with formal parallax uncertainties better than 10\%
were extracted from the Catalog of Galactic OB stars
\citep{ree03}. The spiral structure demonstrated by the {\it Gaia} OB
stars agrees well with that illustrated by the VLBI HMSFR
masers. These OB stars also extend the arm segments traced by HMSFR
masers into the fourth Galactic quadrant.
\citet[][]{chh19} identified 6,858 candidates of O- and early B-type
stars. Together with the known spectroscopically confirmed O-B2 stars
from literature, a sample of 14,880 OB~stars/candidates with {\it
  Gaia} DR2 parallax uncertainties better than 20\% was obtained, and
used to delineate the arm segments in the Solar neighborhood.
Recently, {\it Gaia} published its Early Data Release 3, the parallax
accuracies have been improved significantly, to be
0.02$-$0.07~mas for $G<$17.
In a recent work, \citet[][]{xhb+21} compiled the largest sample of
spectroscopically confirmed O-B2 stars \citep[][]{skiff14} available
to date with astrometric measurements of {\it Gaia} EDR3, including
14,414 O-B2 stars. 9,750 of them have parallax uncertainties better
than 10\%. With this sample, the spiral structure within $\sim$5 kpc
of the Sun are delineated in detail.

In this work, the sample consisting of about 1,090 O-type stars given
by \citet[][]{xhb+21} is adopted. Their distribution in the Galactic
disk is shown in Fig.~\ref{dis}d.
As discussed in \citet[][]{xhb+21}, the distribution of O-type stars
in spiral arms are clumped.
The Sagittarius-Carina Arm traced by O-type stars seems to be not
continuous in the direction of $l\sim$315$^\circ-340^\circ$. A gap of
O-type stars in the range of $l\sim$150$^\circ-160^\circ$ in the
Perseus Arm is also noticed. These properties are consistent with the
results shown by using GMCs and $\hii$ regions.

\subsection{Young open clusters}
An open cluster is a group of stars that formed in a giant molecular
cloud.
In comparison to individual stars, it is possible to estimate more
accurate values of distance, proper motions and radial velocity for an
OC, as it has many member stars.
In our Galaxy, star formation occurs mainly in spiral arms. Hence, the
majority of young OCs are believed to be borned in spiral arms, and
too young to migrate far from their birth locations.
It is accepted that young OCs (e.g. $<$20~Myr) can be used as good
tracers for the nearby spiral arm segments. In comparison, older OCs
have more scattered distribution.
 
\cite{beck63, beck64} first used 156 OCs to study the spiral
structure in the Solar neighborhood. 
They suggested that the distribution of those clusters with earliest
spectral type between O and B2 follows three spiral arm segments.
In comparison, the distribution of the clusters with earliest spectral
type between B3 and F does not present arm-like structures and seems
to be random.
The picture was extended by \cite{beck70} and \cite{fb79}.
Meanwhile, a different point of view raised by \citet{janes82} and
\citet{lynga82} with a large sample of about 400 open clusters. They
suggested that the observed nonuniform distribution of OCs is affected
by the interstellar obscuration, and dominated by the locations of
dust clouds rather than by the spiral structure. \citet{jan88}
mentioned that the young clusters define three complexes, but their
association to a spiral structure is not obvious.
After that, the number of Galactic open clusters gradually increased
\citep[e.g., see][]{mer95,dias02,kp13,sch14,sch15}. Meanwhile, the
spiral structure of the Milky Way was better uncovered by
multi-wavelength observations \citep[e.g.,][]{gg76,ch87,dht01,rus03},
especially in radio band, where the dust obscuration has neglectable
influence on the results.
There has been a general consensus that the nonuniform distribution of
nearby OCs is related to spiral arms
\citep[e.g.][]{dias2005,carr05,Moitinho2006,Vazquez2008,Moitinho2010,cama13,bobylev2014,dias19}

Before the data release of {\it Gaia}, there have been more than 3,000
known OCs \citep[e.g. see,][]{dias02,kp13,dias14,sch14,sch15}. Many of
them have determined mean proper motions and membership
probabilities~\citep[][]{sam17,dias18}. Since the publication of {\it
  Gaia} DR2 \citep[][]{dr2},
on the one hand, the parameters of known OCs have been updated
\citep{can18,sou18,boss19,md19,can20,can20b,tar21,dias21}.
On the other hand, a large number of new OCs and candidates have been
identified
\citep[e.g. see][]{cast18,can19,cast19,sim19,liu19,hao20,cast20,ferr20,he20,hun21,ferr21}.
With OCs, the nearby spiral arms were studied by \citet{can18},
\citet{dias19}, \citet{can20b}, \citet{tar21}, and \citet{ferr21} in
the past few years. Especially, \citet[][]{mon21} studied the spiral
arms traced by young OCs with the updated OC catalogue of
\citet[][]{dias21}, and determined the spiral pattern rotation speed
of the Galaxy, the corotation radius and the statistic properties of
OC parameters. \citet{pog21} studied the spiral structure in the Solar
neighborhood with samples of young upper main sequence stars,
classical Cepheids and open clusters. The open clusters used by
\citet{pog21} is from \citet{can20b}.
Hao et al. (2021) compiled a catalogue of more than 3,700 OCs from the
references above, and re-calculated the parameters (parallaxes, mean
proper motions, radial velocities) based on the latest {\it Gaia} EDR3
\citep[][]{edr3}. The ages of these OCs were either collected from
references or estimated by their analysis.

In this work, we adopted the OC sample provided by Hao et al. (2021),
in which the OC parameters are based on the latest {\it Gaia}
EDR3. There are 627 young OCs (ages $<$ 20~Myr) in their
catalogue. The distribution of young OCs in the Galactic disk is shown
in Fig.~\ref{dis}e.
The distribution of young OCs resembles that of GMCs, HMSFR masers,
$\hii$ regions and O-type stars.
When weighted the OCs with their number of member stars, one can
notice that the OCs with more member stars are more inclined to be
located in spiral arms.

\subsection{Evolved stars}

As the velocity dispersion of gas is smaller than that of the old
stars, the gas response to any perturbations in the stellar disk is
highly amplified \citep[][]{db14}, making the gas arms easier to
identify than the stellar arms traced by old and evolved stars.
For our Milky Way, unlike the gas arms well depicted by GMCs, HMSFR
masers, $\hii$ regions, young OB stars or young OCs, the spiral arms
traced by old stars are still not clear. The properties of stellar
arms are important to better constrain the formation and evolution of
gas arms.

It is commonly suggested that the spiral structure traced by old stars
is dominated by two major spiral arms (the Scutum-Centaurus Arm and
Perseus Arm) based on analyzing the arm tangencies \citep[][]{cbm+09}.
In observations, spiral arm tangencies are indicated by the local
maxima in the integrated number count of old stars or in the
integrated emission in the near-infrared and/or far-infrared bands
against the Galactic longitude, and have only been clearly identified
for the Scutum-Centaurus Arm \citep[][]{dri00,ds01,cbm+09}.
Clear indications of tangencies corresponding to the stellar
Sagittarius-Carina Arm and Norma Arm have not been found from
observational data.
As the Sun is located inside the Perseus Arm, the observed line of
sight do not penetrate its tangency. The method of analyzing tangency
points cannot be applied to this arm.
However, the Perseus Arm is also suggested to be a major stellar arm
based on symmetry. Indeed, many grand-design spiral galaxies with two
well defined spiral arms are observed in the universe
\citep[e.g.][]{galaxyzoo2}.
In comparison, arm tangencies for the gaseous Sagittarius-Carina Arm,
Scutum-Centaurus Arm and Norma Arm could be identified from the
wealthy survey data of radio recombination lines, $\hii$ regions, CO
lines, dense molecular clumps and $\hi$ 21-cm line
etc. \citep[see][]{hh15}.

\begin{figure}
  \centering 
  \includegraphics[width=0.9\textwidth]{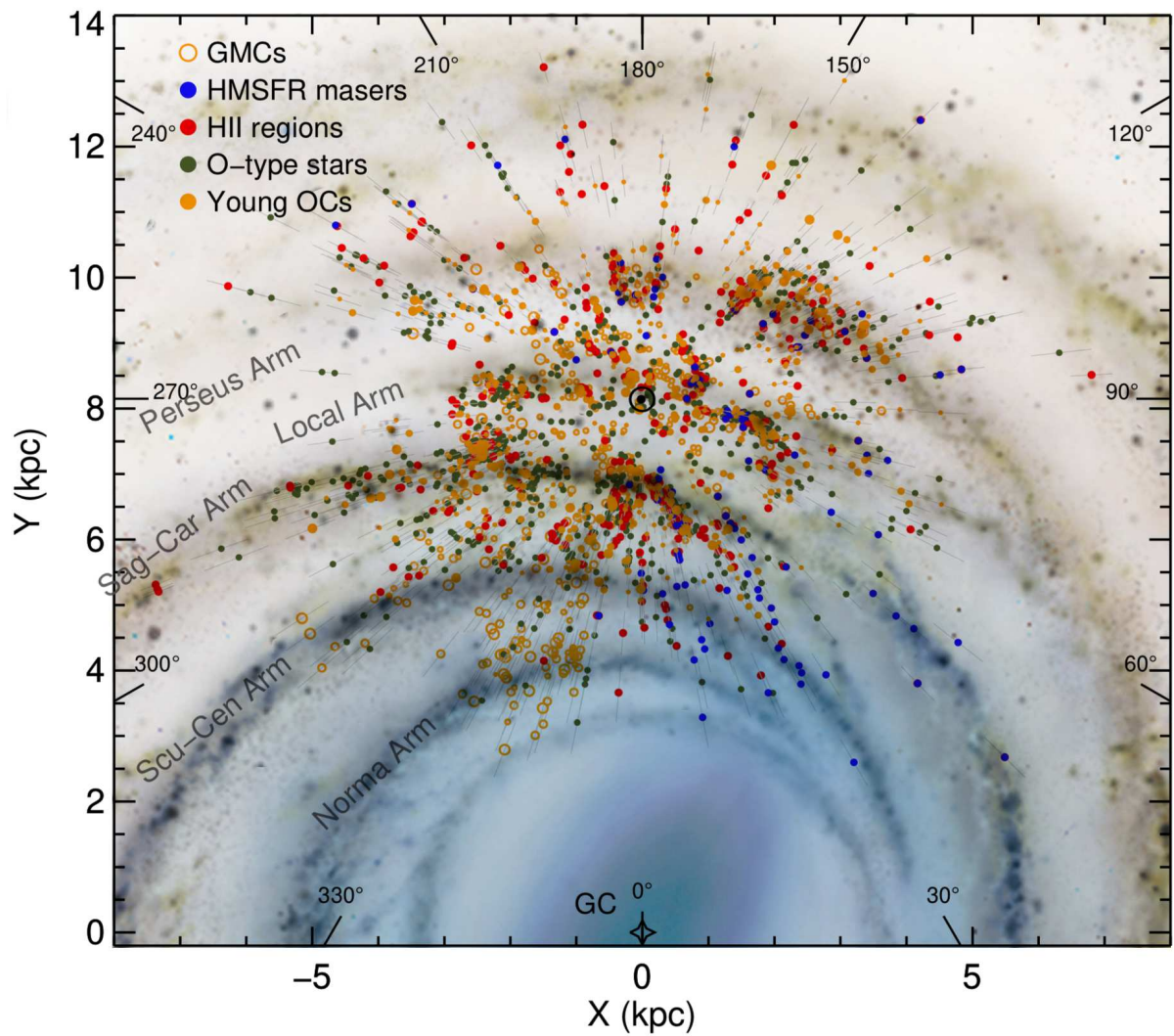}
  \caption{Distributions of the combined dataset of GMCs (yellow
    circles), HMSFR masers (blue dots), $\hii$ regions (red dots),
    O-type stars (green dots) and young OCs (yellow dots) are overlaid
    on a new concept map of the Milky Way, which is credited by:
    Xing-Wu Zheng \& Mark Reid BeSSeL/NJU/CFA based on the spiral arm
    model of \citet{rmb+19}.}
\label{over}
\end{figure}

By taking advantage of the data quality and large sky coverage of {\it
  Gaia}, it is possible to map the nearby spiral structure traced by
evolved stars.
\citet{msk+19} studied the surface density distribution of stars aged
$\sim$1~Gyr, and identified a marginal arm-like overdensity in the
longitude range of $90^\circ \leqslant l \leqslant 190^\circ$. The
overdensity of stars is close to the Local Arm defined by HMSFRs.
By analyzing the {\it Gaia} DR2 data, \citet{kc19} identified a number
of clusters, associations, and moving groups distributed within
$\sim$1~kpc of the Sun, many of them appear to be filamentary or
string-like. The youngest strings ($<$100 Myr) are orthogonal to the
Local Arm. The older ones are suggested to be the remnants of several
other arm-like structures which cannot be traced by dust and gas any
more.
With a new method of analyzing the six-dimensional phase-space data of
{\it Gaia} DR2 sources, \citet{kgd+20} identified six prominent
stellar density structures in the guiding coordinate
space\footnote{The ``guiding coordinate space'' is defined as
  \citep[please see][for a detail]{kgd+20}: $X_g=-R_gsin(\phi)$,
  $Y_g=-R_gcos(\phi)$, here, $R_g=L_z/V_{\rm LSR}$ is the guiding
  radius, $L_z=R \times V_\phi$ is the instantaneous angular momentum
  of the star, $R$ is the Galactocentric distance, $\phi$ is the
  azimuthal angle around the Galactic center clockwise from the
  direction towards the Sun, $V_\phi$ is the azimuthal velocity in the
  Galactic plane.}, corresponding to a physical spatial coverage of
about 5~kpc from the Sun. Four of these structures were suggested to
correspond to the Scutum-Centaurus, Sagittarius, Local, and Perseus
Arms, while the remaining two may be associated with the main
resonances of the Milky Way bar and the outer Lindblad resonance
beyond the Solar circle.
While \citet{hjp+20} presented a different point of view. They
suggested that the stellar density structures identified by
\citet{kgd+20} are known kinematic moving groups, rather than coherent
structure in physical space such as spiral arms.
In addition, it has also been shown that very different bar and spiral
arm models can be tuned to explain the observed features of {\it Gaia}
data \citep[e.g.][]{hbb+19,mfs+19,ehr+20,kgd+20,cfs+21,tfh+21}, making
the situation more difficult to handle.

In the past few years, progress in mapping the stellar arms in the
Solar neighborhood have been made, but there are no conclusive
observational results.
As the properties of nearby stellar arms are not clear, we do not
incorporate them into our analysis/discussions in the following. But
we emphasise that determination of the properties of stellar arms in
our Galaxy definitely deserves more attention.

\begin{figure*}
  \centering 
  \includegraphics[width=0.65\textwidth]{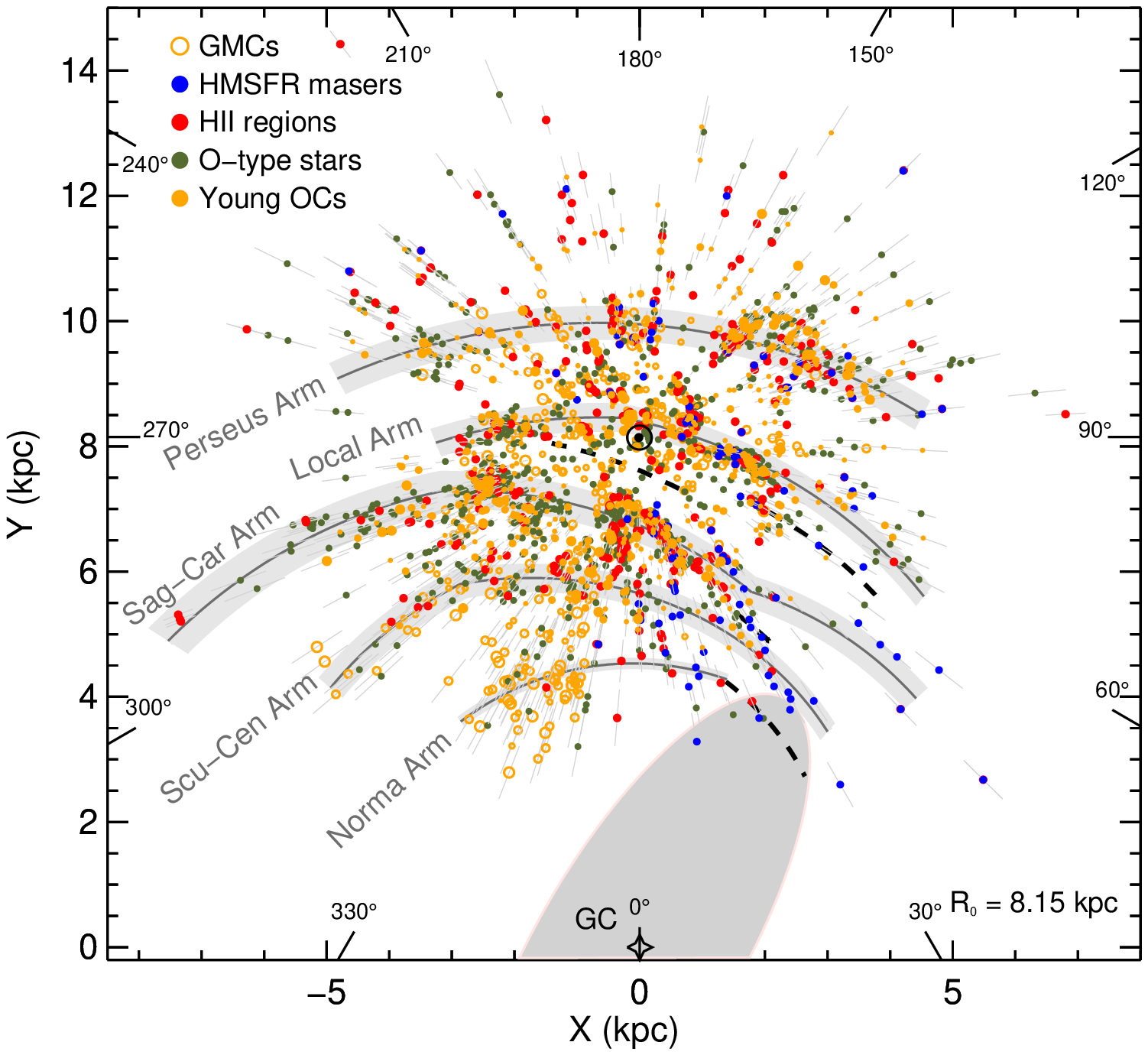} \\
  \includegraphics[width=0.65\textwidth]{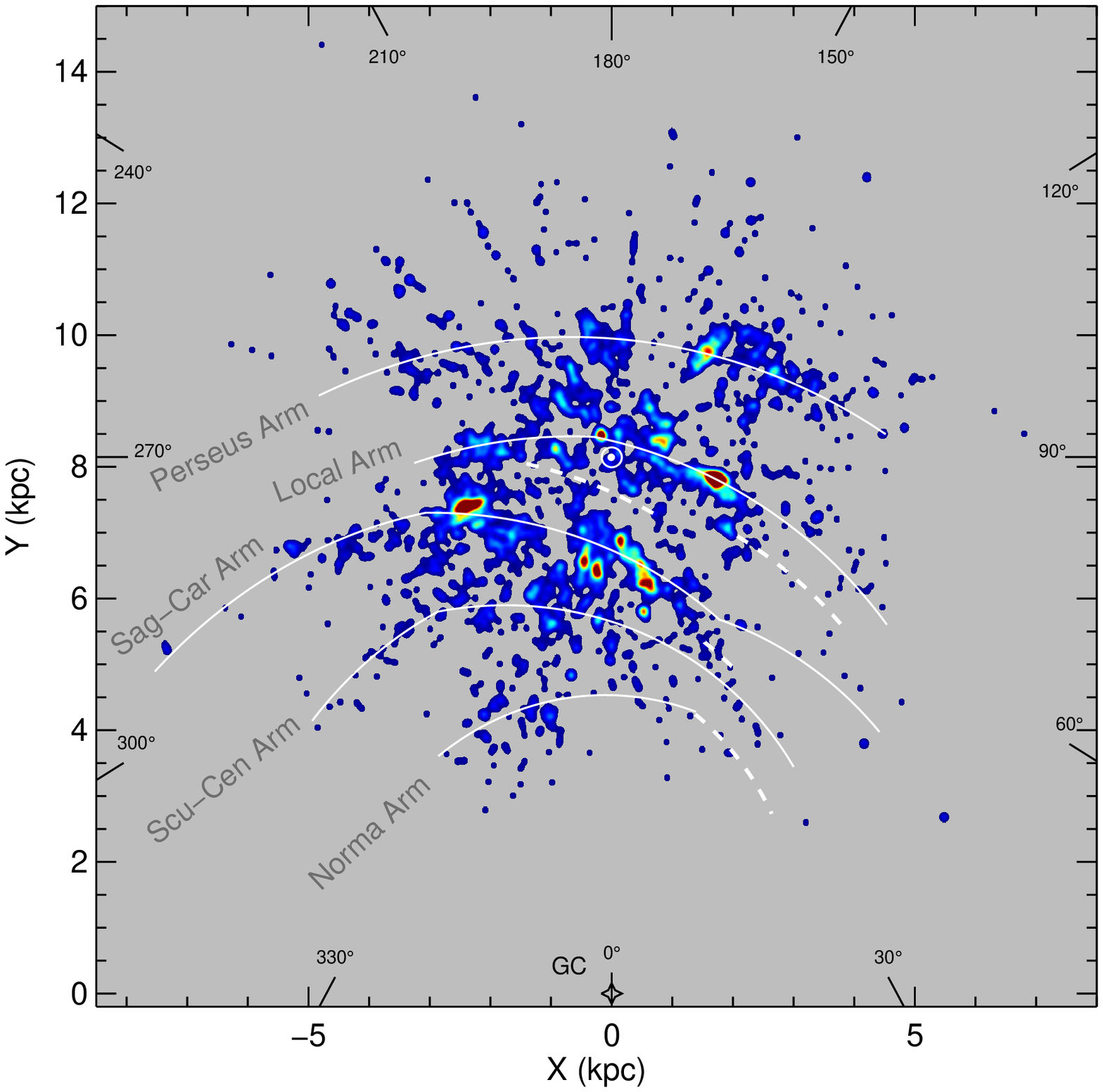}  
  \caption{{\it Upper}: Distributions of the combined dataset of GMCs,
    HMSFR masers, $\hii$ regions, O-type stars and young OCs in the
    Galactic disk. For all the plotted sources, their distance
    uncertainties are better than 15\% and smaller than 0.5~kpc. The
    positional uncertainty for each data point is shown by an
    underlying gray line segment. The curved solid lines indicate the
    best fitted spiral arm model given by this work
    (Table~\ref{para}). The shaded areas around spiral arms denote the
    fitted arm widths. The four dashed lines are the spurs or
    spur-like structures proposed in literature. {\it Lower}: similar
    to the {\it upper} panel, but a ``density'' distribution is
    calculated according to the distribution of the sources.}
\label{disall}
\end{figure*}

\begin{figure*}
  \centering 
  \includegraphics[width=0.98\textwidth]{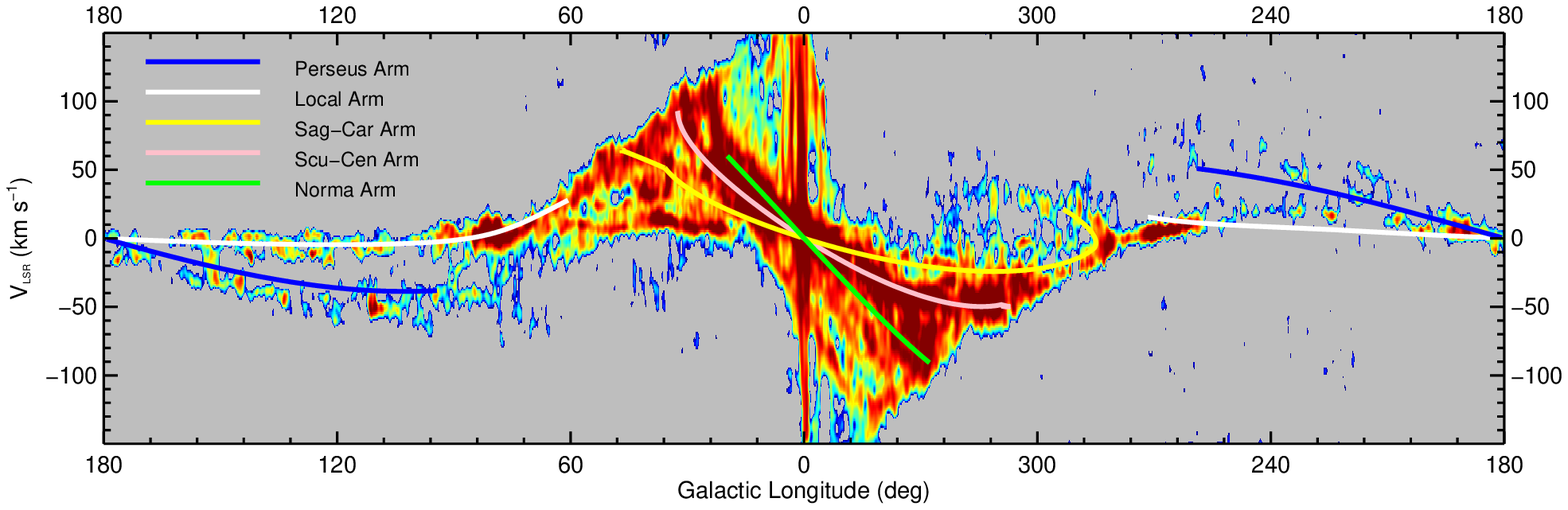} \\ 
  \caption{Best fitted spiral arm model give by this work
    (Table~\ref{para}) are overlaid on the $l-v$ diagram of $^{12}$CO
    (1$-$0) of \citet{dht01}. The spiral arm segments are indicated by
    different colors. To convert the position of spiral arms into the
    $l-v$ diagram, the fitted ``universal'' form of Galaxy rotation
    curve given by \citet{rmb+19} was adopted. The distance of the Sun
    to the GC, $R_0$ is taken as 8.15~kpc. The circular orbital speed
    at the Sun $\Theta_0$ is 236~km~s$^{-1}$ \citep[][]{rmb+19}. }
\label{lvmap}
\end{figure*}

\section{Properties of spiral structure in the Solar neighborhood}

In order to better reveal the properties of spiral structure in the
Solar neighborhood, we combine the data of good tracers of gas arms,
i.e. GMCs, HMSFR masers, $\hii$ regions, O-type stars and young
OCs. It is probably the most wide spread dataset of spiral tracers
with accurate distances available to date. Especially, the data
distributions of different types of tracers are complementary to each
other in the sky coverage. As shown in Fig.~\ref{dis} and
Fig.~\ref{over},
although some substructures seem to exist in spiral arms or inter-arm
regions, the distributions of these objects are in general follow the
dominant spiral arms identified by previous works
\citep[e.g.][]{gg76,rus03,hh14,rmb+19}. Five segments of spiral arms
are delineated by the combined dataset.
They are related to the Perseus, Local, Sagittarius-Carina,
Scutum-Centaurus and Norma Arms from outer Galaxy to the GC direction.
Additionally, there are about 30 sources possibly scattered in the
Outer Arm defined by HMSFR masers. However, the aggregation of sources
is not obvious. We will not discuss the Outer Arm in this work.

\subsection{Fitting model to tracer distributions}

From the combined dataset of spiral tracers (Fig.~\ref{dis} and
Fig.~\ref{over}), some noticeable features are the deviations of
tracer distributions from the modelled spiral arms given by
\citet{rmb+19} in the 3rd and 4th Galactic quadrants, where the HMSFRs
with trigonometric measurements are largely absent.
For instance, many O-type stars and young OCs in the longitude range
of $l\sim210^\circ-260^\circ$ deviate from the modelled Perseus Arm
towards the GC direction.
Similar feature is found for the Local Arm in the longitude range of
$l\sim260^\circ-280^\circ$.
{\bf In the 4th quadrant, many GMCs, $\hii$ regions and O-type stars
  near $l\sim300^\circ-350^\circ$ deviate slightly from the modelled
  Centaurus Arm of \citet{rmb+19} towards the anti-GC direction.}
Hence, with the combined dataset, it would be helpful to update the
parameters of spiral arms in the Solar neighborhood.

The spiral arms observed in spiral galaxies \citep[e.g.][]{sj98,yu18}
and that predicted by the density-wave theory
\citep[e.g.][]{ls64,shu16} are approximately in the form of
logarithmic, which is characterized by a constant pitch angle.
A simple and pure logarithmic form of spiral arm was commonly adopted
in previous works about the Galaxy's spiral structure
\citep[e.g.][]{rus03,hhs09,val08,rmb+09}.
On the other hand, it is found that the spiral arms in galaxies do not
follow logarithmic spirals perfectly, but seem to be kinked in nature
\citep[e.g.][also see the Whirlpool Galaxy
  M\,51]{ken81,kkc11,hr15,daz19}. The observed spiral arms can be
better described by segments of logarithmic form with different pitch
angles.
In theories, tidal interactions can result in noticeable kinks along
spiral arms in a galaxy \citep[][]{db14}. In simulation, \citet{do13}
found that the perturbers can produce segments, and these segments are
joined at kinks to form spiral arm.
For our Milky Way Galaxy, signs of kinked spiral arms were also
noticed~\citep[e.g.][]{tc93,hh14}, as some of the spiral arms cannot
be well fitted by pure logarithmic spirals (e.g. the
Sagittarius-Carina Arm). With pure logarithmic spirals, it is also
difficult to reproduce the observed $l-v$ maps of CO and HI for the
outer Milky Way \citep[e.g.][]{pet14}.
Following \citet{rmb+19}, in this work we adopt a form of kinked
logarithmic spiral to fit an arm, which do not necessarily have a
constant pitch angle.

We allow one or two ``kinks'' in an arm, which means that two or more
segments with different pitch angles are used to describe a single
spiral arm. For the $i$th spiral arm, the form is described as:
\begin{equation}
\ln(R/R_{i,kink})=-(\beta-\beta_{i,kink})\tan\psi_{i},
\label{eq_log}
\end{equation}
here, $R$ is the Galactocentric radius at a Galactocentric azimuth
angle $\beta$. Following \citet{rmb+19}, $\beta$ is defined as
$0^\circ$ toward the Sun and increases in the direction of Galactic
rotation. $R_{i,kink}$ and $\beta_{i,kink}$ are the corresponding
values of $R$ and $\beta$ at the ``kink'' position for the $i$th
arm. $\psi_{i}$ is the pitch angle, which may have an abrupt change at
the ``kink'' position.
To be consistent with \citet{rmb+19} and compare the results with
their models, a Markov chain Monte Carlo (MCMC) approach is adopted in
this work to estimate the arm parameters.
The best-fitted arm parameters are listed in Table~\ref{para},
and the model is shown in Fig.~\ref{disall}. We also compare it with
the observed $l-v$ diagram of CO \citep[][]{dht01} as given in
Fig.~\ref{lvmap}.
In the fitting with the combined dataset, an equal weight is adopted,
although the sky coverage and sample size for different types of
tracers are not the same. Different treatments of weighting parameters
were tested, e.g., scaling according to the sample size of different
types of tracers, but the differences of the fitted arm positions were
found to be small.

In comparison with the best-fitted model of \citet{rmb+19}, the
extension of the Perseus Arm obtained in this work spirals slightly
inward in the third Galactic quadrant. In this work, the Local Arm is
best fit with two segments having different pitch angles, and spirals
inward to the inner Galaxy regions in the third to fourth Galactic
quadrants to match the observational data. The Centaurus Arm in the
fourth quadrant is also fit with two segments in this work. Their
positions are slightly different from those of \citet{rmb+19} {\bf
  near $l\sim300^\circ-350^\circ$} but match the distributions of the
collected GMCs, $\hii$ regions, O-type stars and young OCs. For the
Sagittarius-Carina Arm and the Norma Arm, our fitted arm positions are
in general consistent with that of \citet{rmb+19}.

To evaluate that how well do the different types of tracers fit the
spiral arms, we calculate the percentages of tracers falling into our
best fitted spiral arms. The arm widths given in Table~\ref{para} were
adopted to denote the spiral arm regions. In the Norma Arm, the known
HMSFR masers, $\hii$ regions, O-type stars and young OCs are still
rare. This arm was not used in the calculation. Additionally, the
Outer Arm was also omitted. The percentages of tracers falling into
spiral arms are estimated to be 44\%, 53\%, 58\%, 65\% and 53\% for
the GMCs, HMSFR masers, $\hii$ regions, O-type stars and young OCs,
respectively.
It seems that the GMCs are less confined to the spiral arms than the
other types of tracers. In comparison, if the objects are randomly
distributed in the Solar neighborhood, about 33\% of them are expected
to be in the spiral arms.

\subsection{Spiral arms in the Solar neighborhood}

By taking advantage of the combined dataset of different types of
tracers, segments of spiral arms in the Solar neighborhood have been
delineated and fitted. Now we discuss their properties in detail.

{\bf Perseus Arm}:
The Perseus Arm is a dominant arm in our Galaxy as indicated by
high-mass star-formation activity (HMSFR masers, $\hii$ regions,
O-type stars), young OCs, molecular gas and $\hi$ gas
\citep[e.g.][]{mor53,ch87,rus03,hh14,rmb+19}.
Additionally, the Perseus Arm has been suggested as one of the two
dominant stellar arms of the Milky Way \citep[][]{dri00,ds01,cbm+09}.
As shown in Fig.~\ref{disall}, different types of spiral tracers
(GMCs, HMSFR masers, $\hii$ regions, O-type stars and young OCs) are
mixed together. Their distributions are in general consistent with
each other.
The depicted Perseus Arm is as long as $\sim$12~kpc, starts near (X,Y)
= (4.5, 8)~kpc, and extends to the third Galactic quadrant near (X,Y)
= ($-$6, 9)~kpc.
In this arm, the spiral tracers are not evenly distributed but tend to
cluster. There are two obvious aggregation areas of sources, one is in
the longitude range of $l\sim100^\circ-150^\circ$, the other is in
$l\sim170^\circ-190^\circ$, indicating the active star-formation areas
in the Perseus Arm.
Outside these two regions, some sites of GMCs or star-formation are
scattered, and interspersed with regions showing low star-formation
activity and/or low number density of GMCs.
Outside the segment shown in Fig.~\ref{disall},
the extension of the Perseus Arm to the first or fourth Galactic
quadrant could be explored in the $l-v$ diagram of CO and $\hi$ survey
data, but have not been accurately depicted
\citep[][]{xhw18}. \citet{rmb+19} has suggested that the Perseus Arm
may be not a dominant arm as measured by high-mass star-formation
activity over most of its length.
More spiral tracers with accurate distances are needed in order to
reliably trace this arm to distant Galaxy regions.

{\bf Local Arm}: The Local Arm was thought to be a ``spur'' or
secondary spiral feature for a long time
\citep[e.g.][]{gg76,al97,rus03}, until the density of HMSFRs in the
Local Arm was found to be comparable to that of the Sagittarius Arm
and Persues Arm \citep[][]{xlr+13,xrd+16}. The Local Arm traced by
HMSFR masers stretches for $>$6~kpc, which is larger and more
prominent than previously thought. Hence, it is suggested to be a
dominant arm segment.
As shown in Fig.~\ref{disall}, there are a large number of spiral
tracers (GMCs, HMSFR masers, $\hii$ regions, O-type stars and young
OCs) located in this arm, which present complex substructures.
There are several areas where sources are accumulated. One is near
(X,Y) = (1.5, 7.7)~kpc in the first Galactic quadrant, another is near
(X,Y) = ($-$2, 8.2)~kpc in the fourth quadrant. A filament-like
structure appears in the region from (X,Y)~=~(2, 7.5)~kpc to (X,Y) =
(0.8, 9)~kpc, and spirals outward toward the anti-GC direction with
respect to the fitted arm center. Interestingly, a substructure near
$l\sim100^\circ-150^\circ$ and with $d \sim 0.6$~kpc is indicated by
many GMCs, but without associated HMSFR masers, $\hii$ regions, O-type
stars or young OCs, at least shown by the collected dataset.
For the majority of sources in this arm, their distance uncertainties
are less than 10\%, hence, these substructures are believed to be true
features.
In general, the depicted Local Arm could be as long as $\sim$9~kpc,
starts near (X,Y)~=~(4.5, 6)~kpc, and extends to the third and even
fourth quadrant near (X,Y)~=~($-$3, 8)~kpc.
Outside the delineated segment, the extension of the Local Arm is
still not clear. The Local Arm seems to gradually spiral inward in the
fourth Galactic quadrant, becoming very close to the Carina Arm.
{\bf It is indicating that the Local Arm is possibly an arm branch
  locating between the Perseus Arm and the Sagittarius-Carina Arm.}
More observational data are needed to uncover its nature.

{\bf Sagittarius-Carina Arm:} This arm can be clearly traced by GMCs
\citep[][]{gcbt88} and massive star-formation regions
\citep[][]{rus03,ufm+14,hh14}.
There are a large number of sources in this arm. The
Sagittarius-Carina Arm in Fig.~\ref{disall} starts near (X,Y) = (4.5,
4)~kpc, extends to the fourth quadrant near (X,Y) = ($-$7, 5)~kpc, as
long as $\sim$19~kpc.
It is found that this arm cannot be well fitted by a single pitch
angle model, especially in the longitude range of 20$^\circ-40^\circ$
\citep[e.g.][]{tc93,rus03,hh14,rmb+19}.
Three major aggregation areas of sources are noticed. One is near
$l\sim30^\circ$, showing an elongated structure, which is probably a
true feature as it is not only traced by HMSFR masers, but also by
$\hii$ regions, O-type stars or young OCs.
The other two areas are near $l\sim340^\circ-360^\circ$
and close to the tangent region of the Carina Arm
($l\sim280^\circ-290^\circ$).
The distribution of sources in this arm is well consistent with the
model given by \citet{rmb+19}. Outside the long segment traced in
Fig.~\ref{disall},
the extension of the Sagittarius-Carina Arm could be well delineated
by the distribution of GMCs or $\hii$ regions with kinematic
distances, or indicated by the features shown in the $l-v$ diagrams of
CO and $\hi$.

{\bf Scutum-Centaurus Arm:}
Similar to the Sagittarius-Carina Arm, the Scutum-Centaurus Arm has
also been traced by many GMCs and massive star-formation regions. In
addition, it is also suggested to be one of the two dominant stellar
spiral arms of the Milky Way, as the Centaurus Arm tangent was clearly
shown by the evolved stars surveyed by {\it Spitzer}
\citep[][]{dri00,ds01,cbm+09}.
The collected GMCs, $\hii$ regions, O-type stars and young OCs enrich
the sample of spiral tracers in this arm, especially in the fourth
Galactic quadrant.
As shown in Fig.~\ref{disall}, the spiral tracers in this arm seem to
be more evenly distributed than that of the Perseus Arm, Local Arm and
Sagittarius-Carina Arm.
The traced segment of the Scutum-Centaurus Arm starts near
(X,Y)~=~(2.5, 4)~kpc, and extends to the fourth Galactic quadrant near
(X,Y)~=~($-$5, 4.5)~kpc, as long as $\sim$8~kpc.

{\bf Norma Arm:} This arm is distant ($\gtrsim4$~kpc) from the Sun,
and most likely starts near the near end of the Galactic
bar. Previously, the Normal Arm has not been clearly traced by GMCs or
massive star-formation regions. Mainly because the majority of spiral
tracers possibly associated with this arm only have kinematic
distances.
With the collected dataset, it seems that only a segment of the Norma
Arm can be roughly traced by GMCs in the fourth Galactic quadrant. The
number of HMSFR masers, $\hii$ regions, O-type stars or young OCs
related to this arm is still very limited.

\setlength{\tabcolsep}{5.5mm}
\begin{table*}[!t]
  \caption{Parameters of the best-fitted model of spiral arms (see
    Eq.~\ref{eq_log}) with the combined dataset of GMCs, HMSFR masers,
    $\hii$ regions, O-type stars and young OCs. For the $i$th spiral
    arm, $\beta$ Range in column (2) gives the range of the arm
    segment as shown in Fig.~\ref{disall}, $\beta$ is the
    Galactocentric azimuth angle, which is defined as 0$^\circ$ toward
    the Sun and increases in the Galactic rotation
    direction. $\beta_{kink}$ and $R_{kink}$ in column (3) and (4) are
    the corresponding values of $\beta$ and $R$ at the ``kink''
    position, here $R$ is the Galactocentric radius. The pitch angle
    is $\psi_{<}$ for $\beta < \beta_{kink}$, and $\psi_{>}$ for
    $\beta > \beta_{kink}$. The fitted arm width is listed in column
    (7). The Sagittarius-Carina Arm is consist of three segments,
    hence are divided into two parts and listed below separately.}
\begin{center}
\begin{tabular}{ccccccc}
  \hline
  \hline
  Arm   &  $\beta$ Range & $\beta_{kink}$ & $R_{kink}$  &  $\psi_{<}$  & $\psi_{>}$  & Width \\
        &      (deg)       &  (deg)        & (kpc)      & (deg)       & (deg)      & (kpc)  \\
  (1)   &    (2)         &   (3)         &   (4) &  (5) & (6)  & (7)   \\ 
  \hline
Perseus   & $-28 \to 28$   &  32.7        &  9.57     &  3.9         &   19.9      & 0.26  \\
Local     & $-22 \to 39$   &  $-$2.0       &  8.46     &  4.4         &   12.6      & 0.23  \\
Sagittarius-Carina 1   & $-57 \to 17.5$       &  $-$22.8   &  7.92   & 11.9   &  22.2    & 0.33  \\
Sagittarius-Carina 2   & $17.5 \to 48$      &   17.5    &  5.95   & 21.3  &  0.2    & 0.25  \\
Scutum-Centaurus     & $-50 \to 41$       &  $-$26.3   &  6.47   & $-$0.5   &  16.5    & 0.25  \\
Norma                & $-38 \to 18$       &  18.0     &  4.5    & 1.3   &  19.0    & 0.07  \\
  \hline\hline
\end{tabular}
\end{center}
\label{para}
\end{table*}

\subsection{Substructures in the spiral arms or inter-arm regions}
\label{sec:spur}

Besides the extended structure of major spiral arms which probably
wrap fully around the Milky Way, substructures named as branches,
spurs and features are often observed in spiral galaxies
\citep[][]{el80,db06}. As discussed in \citet{la06} and \citet{db14},
there are no formal definitions of these different types of
substructures in observations and/or numerical simulations.  Different
definitions have been adopted \citep[][]{cls03,db06}.
The initial definitions based on observations are as follows:
(1) arm branches are in general the structures locating between two major
spiral arms, and/or where one arm bifurcates into two, branches may
extend from one arm to another \citep[][]{el80,la06,db14}. 
(2) Spurs are shorter features than branches, and indicated by strings of
star formation sites in the inter-arm regions. They jut out from
spiral arms into the inter-arm regions at larger pitch angles than the
arm itself \citep[][]{wea70,el80}. Two or more spurs are commonly
found to be close or parallel to one another.
(3) Feathers are also short features, but indicated by thin dust lanes
or extinction features that cut across spiral arms and have large
pitch angles. Outside the luminous arms, these extinction features
become mostly undetectable \citep[][]{lyn70}.
Arm branches, spurs and features are typically extend away from the
trailing side of spiral arms \citep[e.g. M\,51,][]{db14}.

As shown in Fig.~\ref{disall}, the objects in spiral arms are not
uniformly distributed, resulting in the patchy and/or bifurcate
appearance of spiral arms. Additionally, about 40\% of the collected
sources distribute in the inter-arm regions, also present some
structural features. These substructures may be related to arm
branches, spurs and/or feathers as found in some nearby face-on spiral
galaxies \citep[e.g. M\,51, M\,101,][]{db14,xhw18}. Our knowledge
about these substructures in our Galaxy is very limited.

In our Galaxy, several spurs and/or spur-like structures have been
classified from observations.
In the direction of $l\sim 90^\circ-210^\circ$, a structural feature
named as Orion spur \citep[e.g. Fig.1 of][]{hum70,sk75,kv79} or
Cepheus spur \citep[][]{pan21} was discussed. This feature is
suggested to be located between the Local Arm and the Perseus Arm,
which may even extend to the first Galactic quadrant
\citep[][]{aas16}.
However, the name of Orion spur was also used to indicate the Local
Arm in some literature \citep[][]{al97,car14,ek14}, which brings up a
question: if the Orion spur discussed in the early days
\citep[e.g.][]{hum70} exists or not?
It will be helpful to reinvestigate this question with modern
observational data. Recently, \citet{xrd+16} identified a spur near
the direction of $l \sim 50^\circ$ traced by five HMSFRs with VLBI
parallax measurements, bridging the Local Arm to the Sagittarius Arm
and having a pitch angle of $\sim$18$^\circ$ \citep[$\sim$13$^\circ$
  given by a recent analysis of][]{rmb+19}. The existence of this spur
is also supported by the CO features shown in the $l-v$ diagram.
By analyzing the distribution and peculiar motions of HMSFR
G352.630$-$1.067 and five O-type stars, a possible spur-like structure
is proposed by \citet{clz+19}, which extends outward from the
Sagittarius Arm.
\citet[][]{rmb+19} mentioned that the Norma Arm in the first Galactic
quadrant displays a spur-like structure, which starts at (X, Y) = (3,
2)~kpc near the end of the Galactic bar and extends to about (X, Y) =
(2, 5) kpc at Galactic azimuth angle of $\sim18^\circ$. This structure
has a large pitch angle of $\sim20^\circ$.
In addition, a spur-like structure bridging the Scutum Arm and the
Sagittarius Arm is also mentioned in \citet[][]{rmb+19}, which is
indicated by the distribution and proper motions of six HMSFRs and
has a large pitch angle of $\sim20^\circ$.
These proposed spurs or spur-like structures are plotted in
Fig.~\ref{dis} and Fig.~\ref{disall}. It seems that some GMCs, $\hii$
regions, O-type stars and young OCs are coincident with these
structures in positions.

Except the Orion spur mentioned in some early literature, the known
spurs or spur-like structures identified in the past few years are
nearly all based on the astrometric data of HMSFR masers. In
comparison to HMSFR masers, the GMCs, $\hii$ regions and especially
O-type stars and young OCs with accurate distances have covered a much
wider Galactic range. Hence, it is expected that more substructures
could be identified.
To identify the substructures in the inter-arm regions or spiral arms,
radial velocities and/or proper motions for the sources would be
helpful, which are still not available for many of them.
Additionally, the properties of these substructures will help us to
better understand the formation mechanisms of Galaxy's spiral
structure. As the formation of these features is different for
different spiral arm models \citep[][]{db14}.

\subsection{Formation mechanisms of Galaxy's spiral structure}

Besides accurately mapping the spiral structure, understanding its
formation mechanism is another difficult issue.
Different mechanisms have been proposed, e.g., the quasi-stationary
density wave theory \citep[][]{ls64,ls66}, localized instabilities,
perturbations, or noise-induced kinematic spirals \citep[][]{sc84},
dynamically tidal interactions \citep[][]{tt72}, or a combination of
some of them \citep[][]{db14}.
Although many efforts have been dedicated to elaborate plausible
hypotheses concerning the origin of the dominant spiral arms of the
Galaxy, it is still not conclusive for now.
One way is to analyse the kinematic properties of stars in the
vicinity of the Sun \citep[e.g.][]{wsb13,fsf14,lws+17,kbc18}. However,
it has been shown that very different bar and spiral arm models can be
tuned to look like the local Gaia data \citep[][]{hbb+19}, or
convincingly explain all observed features at once
\citep[e.g.][]{mfs+19,ehr+20,kgd+20,cfs+21,tfh+21}.
The other method is comparing the relative positions of gas arms and
stellar arms~\citep[e.g.][]{rob69,shu16,dp10,db14,hh15,mgf15,hxh21},
which can be used to verify the predictions of different
theories. Observational evidence for the spatial offsets between the
gas arms and stellar arms have been noticed for the tangent regions
\citep[e.g.][]{hh15}. However, for other regions in the Galactic disk,
it is not clear whether the systematic spatial offsets or age pattern
exist or not \citep[][]{mgf15,val18,hxh21}. More tests based on
observations are needed.

In addition, the properties of the Local Arm make the situation more
complex.
Its existence induces some challenge to the density wave theory
applied to our Galaxy \citep[][]{xlr+13,xrd+16}.
Before 2017, no specific mechanism for the origin of the Local Arm has
been proposed. \citet{lmb+17} first interpreted the Local Arm as an
outcome of the spiral corotation resonance, which traps arm tracers
and the Sun inside it \citep[also see][]{mlb+18}. Their modelled
corotation zone looks consistent with the banana-like structure of the
Local Arm shown by the distributions of GMCs, $\hii$ regions, O-type
stars and young OCs in Fig.~\ref{disall}.
In the context of other mechanisms, e.g. localized instabilities,
perturbations, or noise-induced kinematic spirals, the properties of
the Local Arm may be easily interpreted. The Milky Way has been
suggested to be quite different from a pure grand design spiral, but
probably resemble a multi-armed galaxy M\,101, due to the existence of
the Local Arm and the many possible spurs noticed from observational
data \citep[][]{xhw18}. Typically, the localized instabilities are
associated with flocculent or multi-armed galaxies \citep[][]{db14}.

\section{Conclusions and Discussions}

In this work, the spiral structure in the Solar neighborhood are
discussed based on the largest dataset available to date, which
consists of different types of good spiral tracers. They are GMCs,
HMSFR masers, $\hii$ regions, O-type stars and young OCs. All the
collected data have accurate distances with uncertainties $<$15\% and
$<$0.5~kpc. With the dataset, we update the parameters of spiral arm
segments in the Solar neighborhood, and discuss their properties. The
spiral structure traced by GMCs, HMSFR masers, $\hii$ regions, O-type
stars and young OCs are in general consistent with each other. Five
segments of dominant spiral arms in the Solar neighborhood are
depicted, they are the Perseus, Local, Sagittarius-Carina,
Scutum-Centaurus and Norma Arms. However, the extensions of these arm
segments to distant Galaxy regions have not been reliably traced. In
the spiral arms and inter-arm regions, the distributions of spiral
tracers present complex substructures, which are probably true
features as the distance uncertainties of the tracers are small. At
least five spurs or spur-like features have been identified in the
literature by taking advantage of the astrometric data of HMSFR
masers, but more substructures remain to be uncovered with the updated
dataset of different types of good spiral tracers. In comparison to
the gas arms traced by GMCs and star-formation activity, the
properties of stellar arms indicated by evolved stars are still
inconclusive.

There is significant progress in understanding the Galaxy's spiral
structure in the past few years, which is heavily dependent on the
developments of astrometric observations by the VLBI in radio band and
the {\it Gaia} satellite in optical band.
The VLBI observations have the advantage to measure the spiral tracers
in distant Galaxy regions with high accuracies \citep[as high as
  0.006~mas, typically about $\pm$0.2~mas,][]{rmb+19}, and almost not
affected by dust extinction.
BeSSeL is planned to extend to the southern sky \citep[][]{rmb+19},
which will provide parallax and proper motion measurements for many
HMSFRs in the third and fourth Galactic quadrants, where the data of
such kind of measurements are largely absent at present. In the near
future, the SKA is expected to open a new era for the trigonometric
measurements of a large number of HMSFRs and hence for the
investigations on the Galaxy's global spiral structure.
On the other hand, the {\it Gaia} EDR3 has been released in the end of
2020, the parallax uncertainties have been significantly improved to
be 0.02$-$0.03~mas for $G$ band magnitude less than 15, and 0.07~mas
for $G$~=~17.
The full {\it Gaia} DR3 is expected in 2022. {\it Gaia} is still
committing itself to improve the accuracies of parallaxes and proper
motions for a large number of stars.
Although the stars measured by {\it Gaia} suffered from the dust
extinction, so that distant objects cannot be measured, the {\it Gaia}
data have the advantage to reveal the detailed
structures/substructures and kinematic properties in the Solar
neighborhood, at least for the regions within about 5~kpc of the Sun.
In the Solar neighborhood, the segments of dominant spiral arms have
been well traced as discussed in the main text. However, the
properties of substructures in the spiral arms or inter-arm regions,
and the properties of stellar arms traced by evolved stars are still
far from conclusive, which may be deserving of more attention.

\section*{Acknowledgments}
The author thanks the anonymous referees for constructive comments and
suggestions that significantly improved this work, and Prof. Y. Xu for
helpful suggestions. The author also would like to thank Dr. C. J. Hao
for kindly providing the data of young OCs, and Dr. X. Y. Gao for
carefully reading the manuscript.
This work is supported by the National Key R\&D Program of China
(NO. 2017YFA0402701) and the National Natural Science Foundation
(NNSF) of China No. 11988101, 11933011, 11833009. L.G.H thanks the
support from the Youth Innovation Promotion Association CAS.
This work has made use of data from the European Space Agency (ESA)
mission Gaia (https://www.cosmos.esa.int/gaia), processed by the Gaia
Data Processing and Analysis Consortium (DPAC,
https://www. cosmos.esa.int/web/gaia/dpac/consortium). Funding for the
DPAC has been provided by national institutions, in particular the
institutions participating in the Gaia Multilateral Agreement.

\bibliographystyle{frontiersinSCNS_ENG_HUMS} 
\bibliography{ms}

\end{document}